\documentclass[12pt, reqno] {article}
\usepackage{latexsym}
\usepackage{amsmath}
\usepackage{mathrsfs}
\usepackage{amsfonts}
\usepackage{graphicx}

\usepackage[totalheight = 21cm, totalwidth = 17cm]{geometry}
\newcommand{\bea}{\begin{eqnarray}}
\newcommand{\eea}{\end{eqnarray}}
\newcommand{\be}{\begin{equation}}
\newcommand{\ee}{\end{equation}}
\numberwithin{equation}{section}

\allowdisplaybreaks

\begin{document}
\begin{titlepage}  
\pagestyle{empty}
\baselineskip=21pt
\vspace{2cm}
\rightline{ACT-03-05,CERN-PH-TH/2005-052,MIFP-05-06}
\vspace{1cm}
\begin{center}
{\bf {\large The String Coupling Accelerates the Expansion of the 
Universe}}
\end{center}
\begin{center}
\vskip 0.2in
{\bf John~Ellis}$^1$, {\bf Nikolaos E. Mavromatos}$^{2}$ and 
{\bf Dimitri V. Nanopoulos}$^{3}$
\vskip 0.1in
{\it
$^1${TH Division, Physics Department, CERN, CH-1211 Geneva 23, Switzerland}\\
$^2${Theoretical Physics, Physics Department, 
King's College London, Strand WC2R 2LS, UK}\\
$^3${George P. and Cynthia W. Mitchell Institute for Fundamental 
Physics, Texas A\&M
University,\\ College Station, TX 77843, USA; \\
Astroparticle Physics Group, Houston
Advanced Research Center (HARC),
Mitchell Campus,
Woodlands, TX~77381, USA; \\
Academy of Athens,
Academy of Athens,
Division of Natural Sciences, 28~Panepistimiou Avenue, Athens 10679,
Greece}\\
} 

\vspace{0.5cm}
{\bf Abstract}
\end{center}
\baselineskip=18pt \noindent  

Generic cosmological models in non-critical string theory have a
time-dependent dilaton background at a late epoch. The cosmological
deceleration parameter $q_0$ is given by the square of the string
coupling, $g_s^2$, up to a negative sign. Hence the expansion of the
Universe must accelerate eventually, and the observed value of $q_0$
coresponds to $g_s^2 \sim 0.6$. In this scenario, the string coupling is
asymptotically free at large times, but its present rate of change is
imperceptibly small.

\vspace{1cm}

\leftline{CERN-PH-TH/2005-052}
\leftline{March 2005}
\end{titlepage}
\baselineskip=18pt

\section{Introduction} 

One hundred years after Einstein's `annus mirabilis', theoretical
physicists are struggling to come to terms with the accelerating expansion
of the Universe that is indicated by recent 
cosmological observations~\cite{snIa,wmap}.
What Einstein once termed his `greatest blunder', namely a possible
cosmological constant, may actually turn out to be one of his deepest
insights and a puzzle for quantum theories of gravity~\cite{carroll}. 
This vacuum energy is a second measurable quantity that, in
combination with the Newton constant $G_N \equiv 1/m_P^2: m_P \sim
10^{19}$~GeV, must be confronted with any theory of gravity. Specifically,
the cosmological constant $\Lambda \sim 10^{-48}$~GeV$^4$ is a challenge
for any candidate quantum theory of gravity, which must explain not only
why it is non-zero, but also why it is so many orders of magnitude smaller
than the apparently natural order of magnitude $\Lambda \sim m_P^4$.

This challenge is acute for string theory~\cite{gsw,polchinski},
particularly in its standard paradigm as a conformal field theory on an
internal world sheet, used to calculate an S-matrix for particle
scattering in a static background. The central problem is that a Universe
with a cosmological constant is described as a de Sitter space. This
possesses an event horizon and requires a description of physics in terms
of mixed quantum-mechanical states, which does not admit an S-matrix
formulation of scattering~\cite{smatrix}.

One of the first attempts to transcend the standard paradigm of a 
static string background was a formulation of string in a 
time-dependent dilaton background~\cite{aben}. 
This model can be interpreted as a 
non-critical string~\cite{ddk}, 
in which the underlying effective field theory on 
the internal world sheet is no longer conformal. The deviation from 
conformal symmetry requires the introduction of a renormalization 
scale, which can be interpreted as a new scalar Liouville field in the 
effective world-sheet theory.

We have argued~\cite{emn} that the zero mode of the Liouville field in
such a non-critical string theory can be identified with time, as dictated
in some explicit examples~\cite{gravanis} by the energetics of the
corresponding effective field theory in space-time. One of the miracles of
standard critical string theory was to derive Lorentz invariance and hence
Einstein's Special Relativity. Conversely, one possible signature of
non-critical string could be a deviation from Lorentz
invariance~\cite{aemn}, and we have suggested that distant astrophysical
sources of energetic photons could provide sensitive probes of this
possibility~\cite{nature}.

Here we further argue that any deviation from criticality in the string 
world-sheet theory can be regarded effectively as vacuum energy in 
four-dimensional space-time. 
This could provide a stringy framework for discussing cosmological 
inflation~\cite{brany,EMNW}. 
However, here we focus on another possible application of 
this idea, namely as a mechanism for generating the present-day vacuum 
energy~\cite{EMNDE}. 
In a wide class of models, this suggestion has a startling 
implication that we would like to emphasize. The cosmological 
deceleration parameter $q_0$, which is directly related to the vacuum 
energy, can be expressed in terms of the dilaton field value, which can 
in turn be identified with the string coupling strength. Therefore, the 
cosmological deceleration at asymptotically late times is a direct 
measure of the string coupling strength:
\begin{equation}
q_0 \; = \; - g_s^2.
\label{basic}
\end{equation}
This remarkable equation is our central result. We do not know whether 
the present Universe is sufficiently asymptotic for the formula 
(\ref{basic}) to be directly applicable to the present cosmological 
data. However, we do note one fundamental implication of (\ref{basic}): 
because of its negative relative sign, it implies that {\it the expansion 
of the Universe MUST accelerate eventually}. Moreover, if we insert the 
present measurement of $q_0$, we estimate $ g_s^2 \sim 0.6$, which is 
quite a acceptable value. The string coupling decreases towards zero at 
large time, but the present rate of change is very small.

\section{Background Analysis} 

We now explain in more detail~\cite{EMNDE} the theoretical analysis 
leading to (\ref{basic}). The Ansatz of~\cite{aben} for a string model 
of cosmology was that the dilaton field $\Phi$ could evolve linearly in 
the world-sheet time variable $t$:
\begin{equation}
\Phi \; = \; {\rm constant} - Q t,
\label{linear}
\end{equation}
where $Q$ is a constant whose square measures the departure of the 
world-sheet field theory from conformal symmetry. Since the Einstein 
term in the effective space-time action is conformally rescaled by a 
factor $e^{-\Phi}$, the cosmological time $t_E$ in the Einstein frame 
(in which the 
lowest-order curvature term in the target-space effective action has the 
same normalisation as the conventional Einstein scalar curvature 
term~\cite{gsw}) is related to $t$ by
\begin{equation}
t_E \; = \; c_1 + {c_0 \over Q} e^{Qt},
\label{tEt}
\end{equation}
and the resulting form of a spherically-symmetric four-dimensional 
metric is of the flat Robertson-Walker-Friedman type :
\begin{equation}
ds^2 \; = \; - dt_E^2 + a_E(t_E)^2 (dr^2 + r^2 d\Omega^2),
\label{RWF}
\end{equation}
where $a_E(t_E)$ is a time-dependent scale factor and $\Omega$ is the  
three-dimensional angular factor. 

Such time-dependent cosmological backgrounds are in general described by 
non-critical string theories~\cite{ddk,emn}, in 
which conformal symmetry is restored by 
dressing the field operators with the Liouville 
field, which characterizes the overall size of the string and acts as a 
renormalization scale. The identification of the zero mode of this 
Liouville field with physical time has been checked by many calculations 
from two-dimensional black holes~\cite{emn} to effective potentials between 
D-branes~\cite{recoil,EMNW}.

Non-critical string models generically relax at large cosmic times to 
equilibrium points in the string `theory space', which are conformal field 
theories describing locally-Minkowski spaces with a linear dilaton 
background. One example was a ten-dimensional Type-0 
string theory~\cite{type0} 
compactified on a space with five flat dimensions and a 
non-trivial flux parallel to the remaining dimension~\cite{dgmpp}. 
In this example, the 
sizes of the extra dimensions rapidly froze to fixed values, while the 
positive central-charge deficit $Q^2 > 0$ relaxes to 
a constant value $Q_0^2$ at large times.
Other examples are provided by colliding brane 
worlds~\cite{ekpyrotic,gravanis}.

In such models, the 
scale factor takes the following form at large cosmic times $t_E$:
\begin{equation} 
a_E(t_E) \simeq \frac{\beta Q_0}{\gamma}\sqrt{1 + \gamma^2 
t_E^2}.
\label{aE}
\end{equation}
where $\beta$ and $\gamma$ are numerical constants characteristic
of the specific model under consideration.
For instance, for the type-0 model of \cite{dgmpp}, 
we have $\gamma \equiv (\beta^2 Q_0^2) / (\alpha A)$, where $\alpha = 
\sqrt{[11+\sqrt{17}]/[2(3+\sqrt{17})]}$, $\beta = 2/(1+\sqrt{17})$,
and $A$ is the flux in the large extra dimension.

We see from (\ref{aE}) that $a(t_E)$ scales linearly with $t_E$ at very 
large values of this
Einstein-frame cosmological time~\cite{dgmpp}. Hence the cosmic 
horizon expands logarithmically, allowing for the
proper definition of asymptotic states and thus a scattering matrix.
At large $t_E$, the Hubble parameter becomes 
\begin{equation}
\label{hubble2} 
H(t_E) \simeq \frac{\gamma^2 t_E}{1 + \gamma^2 t_E^2}~,
\end{equation}
and the effective four-dimensional 
vacuum energy density is~\cite{dgmpp}:
\begin{equation} 
\Lambda_E (t_E) 
\simeq \frac{\gamma^2}{\beta^2 ( 1 + \gamma^2 t_E^2)},
\label{cosmoconst} 
\end{equation}
where we use the fact that~\cite{dgmpp} the central-charge deficit
approaches its equilibrium value 
$Q_0$ for large $t_E$. Thus, the dark energy 
density relaxes to zero for $t_E \to \infty$ in such non-critical
string cosmologies~\cite{dgmpp,EMNDE}.
Finally, during this epoch
the deceleration parameter becomes~\cite{EMNDE}:
\begin{equation} 
q(t_E) = -\frac{(d^2a_E/dt_E^2)~a_E}{({da_E/dt_E})^2} 
\simeq -\frac{1}{\gamma^2 t_E^2}.
\label{decel4}
\end{equation}
Therefore, up to proportionality constant factors which by 
convention are normalized to unity, 
it can be identified with the 
square of the string coupling:
\begin{equation}
q(t_E) = - {\rm exp}[2(\Phi - {\rm const})] = - g_s^2.
\label{important}
\end{equation}
which is our central result announced earlier~\footnote{Consistency with 
perturbation theory requires $g_s < 1$, which is easily satisfied in 
phenomenologically realistic string models~\cite{dgmpp,EMNDE}. We note 
also that the present rate of change of $g_s$ is unobservably slow.}. 
Because of the minus sign in 
(\ref{important}), this non-critical string theory predicts that the 
expansion of the Universe must accelerate asymptotically.

\section{Time for Discussion} 

The important ingredient in this approach is the treatment of time as a
dynamical world-sheet renormalisation-group scale~\cite{emn}, which flows
irreversibly between fixed points in string theory space that correspond
to equilibrium theories. The irreversible evolution of this world-sheet
scale is due to information loss associated with world-sheet modes whose
two-dimensional momentum scales pass beyond the ultraviolet cutoff,
leading in turn to microscopic irreversibility of time. Deviations from
such fixed points arise from relevant perturbations that might be due to
catastrophic cosmic events such as the collision of two brane worlds, or
simple quantum fluctuations.  During the irreversible flow to some final
fixed point in the string landscape, the Universe expands and may pass
through various transitions such as inflation and reheating.

As we have just showed, in this scenario the expansion of the Universe
must accelerate, and its rate of acceleration is equal to the current
value of the string coupling. This non-critical string approach predicts a
new type of asymptotic freedom, as the string coupling decreases with
increasing cosmic time.

We close by recalling that this approach to target time in non-critical
string theory yields a number of important, physically falsifiable
predictions. These include potential violations of Lorentz symmetry and
the principle of equivalence, associated with the microscopic curvature of
space-time~\cite{emn,EMNW}, and the possibility that microscopic quantum
mechanics may be modified. One can also use non-critical strings to
discuss supersymmetry breaking in brane worlds~\cite{gravanis} as well as
inflation and reheating~\cite{brany}.  The relative separations and
velocities of recoiling branes in some ekpyrotic models of
inflation~\cite{brany} can be constrained by available and future
astrophysical data, such as Cosmic Microwave background fluctuation
measurements. Thus, a century after Einstein's Special Theory of
Relativity, one scenario for its quantum-gravitational counterpart is
close to experimental test - and possible disproof. {\it Time will tell}.

\section*{Acknowledgements}

N.E.M. wishes to thank Juan Fuster and IFIC-University of Valencia 
(Spain) for their interest and support. 
The work of D.V.N. is supported by D.O.E. grant
DE-FG03-95-ER-40917.

\end{document}